\documentclass[final,5p,times,twocolumn]{elsarticle}

\usepackage{amssymb}
\usepackage{amsmath}
\usepackage{graphicx}
\usepackage{subcaption}
\usepackage{lineno}
\usepackage{xcolor}

\begin{document}

\begin{frontmatter}



\title{Experimental and molecular dynamics study of the
  ionic conductivity in aqueous LiCl electrolytes}


\author{Are Yll\"o}
\author{Chao Zhang\corref{cor1}}
\cortext[cor1]{chao.zhang@kemi.uu.se}
\address{Department of Chemistry-\AA ngstr\"om Laboratory, Uppsala, University, L\"agerhyddsv\"agen 1, 75121 Uppsala, Sweden}

\begin{abstract}

Lithium chloride LiCl is widely used as a prototype system to study the strongly
dissociated 1-1 electrolyte solution. Here, we combined experimental
measurements and classical molecular dynamics simulations to study the ion
conduction in this system. Ionic conductivities were reported at both 20$^\circ$C
and 50$^\circ$C from experiments and compared to results from molecular
dynamics simulations. The main finding of this work is that
transference numbers of Li$^+$ and Cl$^-$ become comparable at high
concentration. This phenomenon is independent of the force fields
employed in the simulation and may be resulted from the ion-specific
concentration dependence of mobility. 

\end{abstract}

\begin{keyword}

Electrolyte solution \sep Ionic conductivity \sep Molecular dynamics
\sep Force fields \sep Debye-Onsager theory   



\end{keyword}

\end{frontmatter}


\section{Introduction}
\label{}

Aqueous electrolytes play important roles in many areas of science and
engineering, such as electrophysiology, electrochemistry and colloid
science. Simple 1-1 electrolyte which is completely dissociated in
dilute solution is often used as a prototype system to develop
analytical theories such the well-known Debye-H\"uckel
theory~\cite{Fawcett:2004ww}. This tradition dates back to the beginning of Physical Chemistry and coins the early physical
chemists as ``Ionists''~\cite{Servos:1996tl}. 

Lithium chloride (LiCl) as an example of these simple 1-1 electrolytes is of particular
interest due to its very high solubility ($\sim$ 45 wt\% at room temperature). The structure of
LiCl solution has been extensively investigated by X-ray diffraction and
neutron scattering experiments ~\cite{Tromp:1992hr, Winkel:2011kn,
  Ansell:2006dv} in together with reverse Monte Carlo  and molecular
dynamics simulations~\cite{Ibuki:2009ca,Harsanyi:2005ju, Petit:2008tw,
  Ibuki:2009ca,  Harsanyi:2011ea, Pluharova:2013jo, Aragones:2014ki, Pethes:2017fr}. The synergy between experiments and simulations
has been proven to be useful to gain a deeper understanding of
solvation structures of Li$^+$ and Cl$^-$.

In the molecular dynamics simulation community, another interest of
modeling LiCl solution was on developing various kinds of force-fields
where cations and anions are commonly described by Lennard-Jones (LJ)
potential and point charge~\cite{ Joung2008, Li:2015gp}. Despite of its simplicity, this approach
has been shown be capable to capture both single ion properties (such
as the hydration free energy) to ion-ion interactions as reflected in
radial distribution functions and the solubility~\cite{Aragones:2014ki}. We refer interested readers to a
recent work on this topic for a comprehensive overview and benchmarks~\cite{Pethes:2017fr}. 

On the other hand, the dynamical and transport properties of these
models were often overlooked. In particular, the ionic conductivity of
LiCl calculated from molecular dynamics simulations has not been
compared to experimental measurements at both room temperature and
elevated temperature. This fact is somehow surprising, because the basic function of any
electrolyte is to serve as an ionic conductor. 

In this work, we carried out both experimental measurements and molecular
dynamics simulations of the ionic conductivity in LiCl solutions.  Ionic
conductivities were reported at both 20$^\circ$C and 50$^\circ$C from
experiments and compared to those calculated from molecular
dynamics simulations using three different force-field models
~\cite{Pluharova:2013jo, Joung2008, Li:2015gp} (See Section 2
for details) and SPC/E water~\cite{Berendsen:1987uu}. In addition to provide reference data for future
force-field developing works, the main finding of our study is that
transference numbers (i.e. the fractional contribution to the ionic
conductivity) of Li$^+$ and Cl$^-$ become comparable at high
concentration. This phenomenon is independent of the force fields
employed in the simulation and can be explained by taking into account
the ion-paring and ion-specific effects. The later imposes a challenge
to the Debye-Onsager theory of the ionic conductivity.

\section{Experimental and computational methods}

\subsection{Ionic conductivity measurements}

The conductivity measurement of LiCl at 2, 5, 10, 15,
  20, 25, 30, 35 and 40 wt \% were performed with an "InLab" conductivity meter (Mettler
Toledo). The conductivity meter probe used is a 4 pole InLab 738-ISM
by (Mettler Toledo) which has a sensitivity range from 0.01--1000 mS/cm and gives accurate measurements up
to 100$^\circ$C. Before measuring, the probe was calibrated
with a standardized 12.88 mS/cm potassium chloride (KCl) solution (Mettler Toledo). After the
successful calibration of the instrument, the probe was lowered into
respective solution. The measurement ran until both the conductivity
and the temperature of the solution had equilibrated at a stable
value. The mean of the
five independent measurements were then noted as the final conductivity of that
solution at 20$^\circ$C.

Similar measurements were then done at an elevated temperature of
approximately 50$^\circ$C. The solutions were heated to 50$^\circ$C by
placing them in a heated water bath with an external thermometer
attached to a reference plastic container with deionized water. When
the solution had reached the sought-after temperature, the
measurements were carried out in the same way as before.
 
\subsection{Molecular dynamics simulations}

\begin{table*}[h]
       \centering
	\caption{Three ion models used in this work.}
	\label{table:potential}
	\begin{tabular}{ c c c c c c }
		\hline
		Model & $\sigma_{\textrm{Li,Li}} $ (nm)&
                                                         $\epsilon_{\textrm{Li,Li}}
                                                         $ (kJ/mol)&
                                                                     $\sigma_{\textrm{Cl,Cl}}
                                                                     $
                                                                     (nm)
          & $\epsilon_{\textrm{Cl,Cl}} $ (kJ/mol) &
                                                    $q_{\textrm{Li}}~/
                                                    q_{\textrm{Cl}}$ (e) \\ \hline
		JC-S \cite{ Joung2008} & 0.1409 & 1.4089 & 0.4830 &
                                                                    0.0535 & $+1~/-1$ \\ 
		LI-IOD-S \cite{Li:2015gp} & 0.2343 & 0.0249 & 0.3852 &
                                                                       2.2240 & $+1~/-1$ \\ 
		PL \cite{Pluharova:2013jo} & 0.1800 & 0.0765 & 0.4100
          & 0.4928 & $+0.75~/-0.75$ \\ \hline
	\end{tabular}  
\end{table*}

The initial cubic box containing simple point charge/extended (SPC/E) water molecules~\cite{Berendsen:1987uu} and random
distributed Li$^+$/Cl$^-$ ions was 2.963 nm for each side.  Water
molecules were kept rigid using the SETTLE algorithm~\cite{Miyamoto1992}. The Ewald
summation was implemented using the Particle Mesh Ewald (PME)
~\cite{Ewald} scheme
and short-range cutoffs for the van der Waals and Coulomb interaction
in the direct space are 1 nm.

Three force fields (ion models) for LiCl were chosen in this study which are
Joung-Cheatham III (JC-S) \cite{ Joung2008},
Li-Song-Merz (LI-IOD-S)
\cite{Li:2015gp} and Pluha\v rov\' a -Mason-Jungwirth (PL)
\cite{Pluharova:2013jo}. JC-S was parameterized against thermodynamic
data such hydration free energy and lattice energy of salt crystal and
has been validated for higher salt concentration~\cite{Zhang:2010zh,
  Zhang:2012fo} at room temperature. LI-IOD-S focus on the structural aspect and was fitted to the ion oxygen distance in
the first solvation shell. PL-S was tuned by scaling down the
point charge of each ion by the refractive index of liquid water in
order to make up the missing electronic polarization. The corresponding LJ
parameters and point charges of these three models are summarized in Table
\ref{table:potential}. In all cases, the Lorentz-Berthelot
combination rule was used between two dissimilar non-bonded atoms.  

Regarding the technical setting in simulations,  the steepest descent
algorithm was used for the energy minimization before the
equilibration. The NVT (constant number of particles, constant volume and constant temperature) equilibration ran for 1 ns with the timestep of 2 fs. The temperature
was then held in place using the Bussi-Donadio-Parrinello thermostat
which preserves both thermodynamic and dynamic properties~\cite{Bussi:2008wu}. The follow-up NPT
(constant number of particles, constant pressure and constant
temperature) simulations ran for 10 ns each and trajectories were
collected every 0.5ps for
conductivity calculation and structural analysis. During the NPT simulations, Parrinello-Rahman barostat~\cite{M.1980} was employed with a
reference pressure of 1.0 bar. This simulation protocol was used for
LiCl solution at 2, 5, 10, 15, 20, 25, 30, 35 and 40 wt \%  and both
20$^\circ$C and 50$^\circ$C and all simulations were performed using
GROMACS 4 package~\cite{Hess2008}. This corresponds to simulations
with following compositions in terms of number ratio $N_\text{salt}$/$N_\text{water}$: 6/694, 15/676,
30/646, 46/614, 61/584, 77/552, 94/518, 110/486, 127/452. Statistical errors were estimated using the standard deviation of observables from 5 equispaced segments in the full trajectory.

\section {Results and discussion}

\subsection{Ionic conductivity and transference number}

The simplest way to calculate the ionic conductivity in molecular
dynamics simulations is to use the Nernst-Einstein equation~\cite{Nitzan:2013tk}:

\begin{eqnarray}
  \sigma &=& \sigma_+ + \sigma_- \\
  & = &\frac{q_+^{2}\rho D_+}{kT} +\frac{q_-^{2}\rho D_-}{kT} 
\label{eq:nersteinstein}
\end{eqnarray}
where $\sigma$ is the ionic conductivity of the solution, $\sigma_+$
and $\sigma_-$ are ionic conductivities for cation and anion
respectively. $q_+$ and $q_-$ are point charges of ions in the
model. $\rho$ is the number density of the salt, $k$ is
Boltzmann constant and $T$ is the temperature.

$D_+$ and $D_-$ are self-diffusion coefficients of cation and anion respectively and
computed from the corresponding mean squared displacement using the
Einstein relation as follows:

\begin{equation}
\label{msd}
  D_{+/-} = \lim_{t\rightarrow \infty}\frac{1}{6t}\frac{1}{N_\text{salt}}\sum_i^{N_\text{salt}}\langle[\mathbf{r}_{i,+/-}(t)-\mathbf{r}_{i,+/-}(0)]^2 \rangle
\end{equation}  
where $t$ is the time, $N_\text{salt}$ is the number of LiCl salt,
$\mathbf{r}_{i, +/-}$ is the position of $i$th cation or anion, $\langle \cdots
\rangle $ indicates the ensemble average. 

One should note that the the Nernst-Einstein equation holds only for
non-interacting charged particles in a homogeneous and isotropic
solvent. Thus, the ion-ion correlation is not taken into account in
the formula. In other words, the ionic conductivity calculated using
the Nernst-Einstein equation gives an upper bound of the actual
value. 

On the other hand, since $\sigma$ is a sum of individual contributions
of cations and anions by construction, the transference number $t_{+/-}$ can be readily extracted
as:

\begin{equation}
  t_{+/-} = \frac{\sigma_{+/-}}{\sigma}
\end{equation}

\begin{figure}[h]
  \centering
  \includegraphics[width=\linewidth]{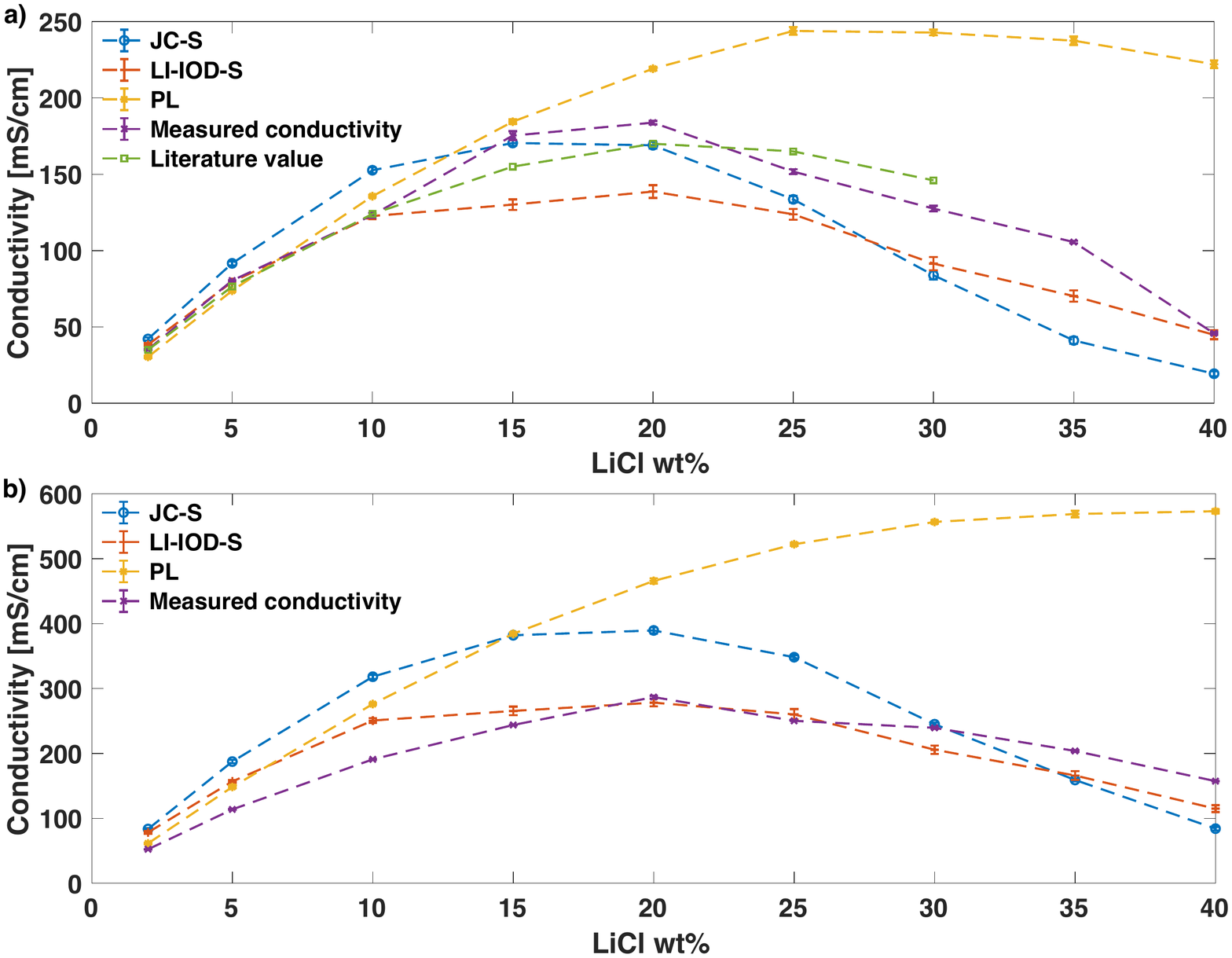}
  \caption{Ionic conductivities vs. wt\% of LiCl from MD
    simulations and experimental measurements at 20$^\circ$C a) and
    50$^\circ$C b). Literature value at 20$^\circ$C is from
    Ref.~\cite{CRC:99}.} \label{fig:mdcond}
\end{figure}

\begin{figure}[h]
	\centering
        \includegraphics[width=\linewidth]{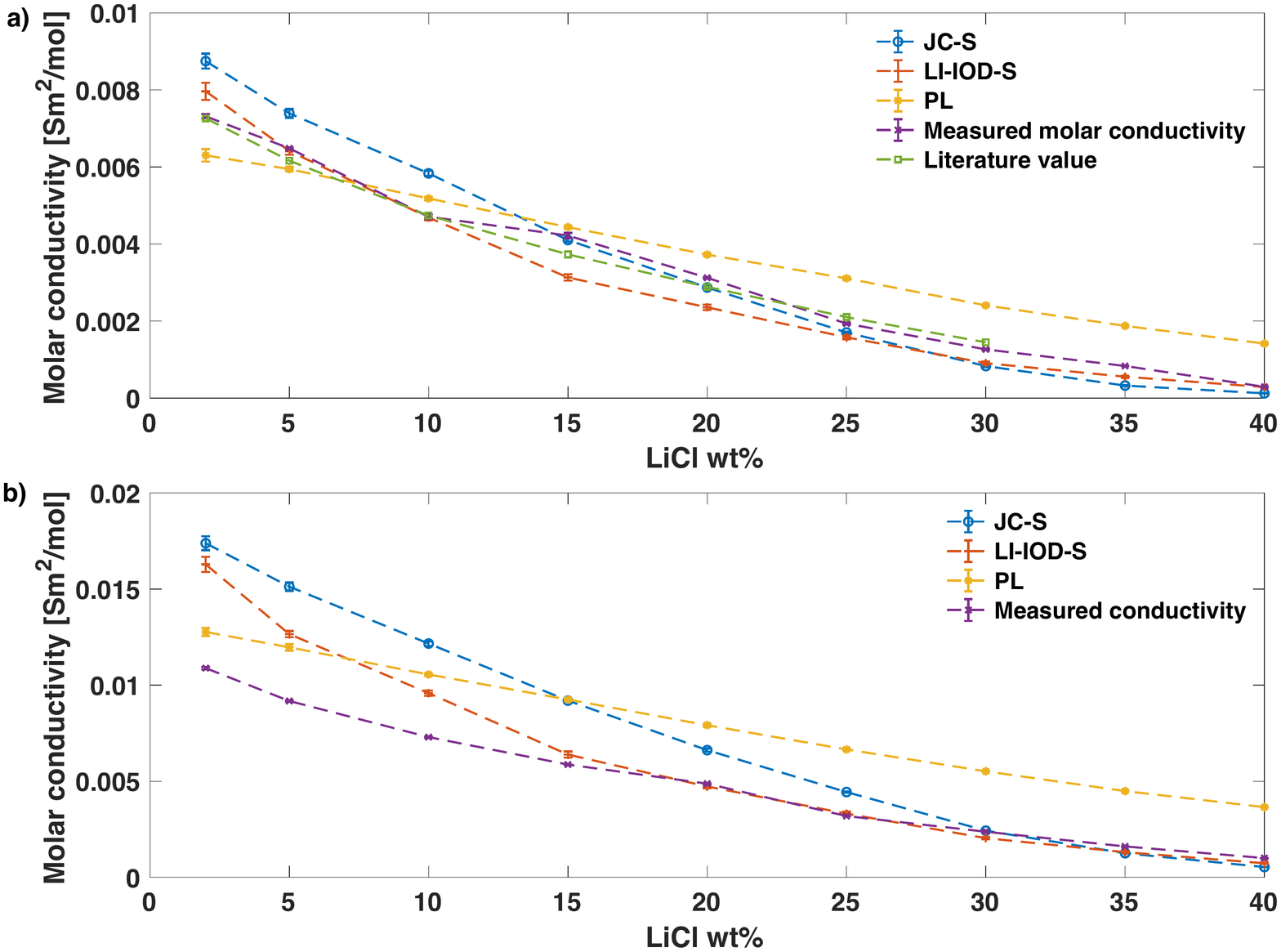}
	\caption{Molar conductivities vs. wt \% of LiCl from MD
          simulations and experimental measurements at 20$^\circ$C a)
          and 50$^\circ$C b).}
	\label{fig:mdmolcond}
\end{figure}

From Fig.~\ref{fig:mdcond}a, we see that the results from  JC-S is
the one that comes closest to the measured and the literature values
in the whole concentration range at 20$^\circ$C, although three ion
models seem be equally well at lower concentrations. Our measured
molar conductivity is also in accord with the result in a recent report~\cite{Yim:2018gq}. At 50$^\circ$C
(See Fig.~\ref{fig:mdcond}b),
LI-IOD-S gives results which agrees best with measured values. JC-S
overestimates the conductivity for lower to mid-range concentrations
and underestimates it for higher concentrations. At both temperatures, PL
significantly overestimates the conductivity from mid to high
concentrations. Similar behavior of PL has been reported for the diffusion
coefficient of Li$^+$ and Cl$^-$ recently~\cite{Pethes:2017fr}. This is likely due to the fact that point charge
of ions are scaled down in this model which leads to a much weaker
ion-solvent interaction.

Both JC-S and LI-IOD-S manage to describe the parabola behavior of the
ionic conductivity as a function of the concentration and to provide
accurate estimates of the corresponding concentration at the conductivity
maximum. The reason for the conductivity maximum comes from a tradeoff
between the increase of number of charge transportors and the decrease of their
mobility as the concentration goes up.  When molar conductivities are
plot instead (Fig.~\ref{fig:mdmolcond}), one can see clearly that the mobility of ions reduces as
a function of the concentration.  Results of JC-S and LI-IOD-S have
better agreements with experiments while PL shows a much higher deviation in the mid-to-high concentration range. 

\begin{figure}[h]
	\centering
		\includegraphics[width=\linewidth]{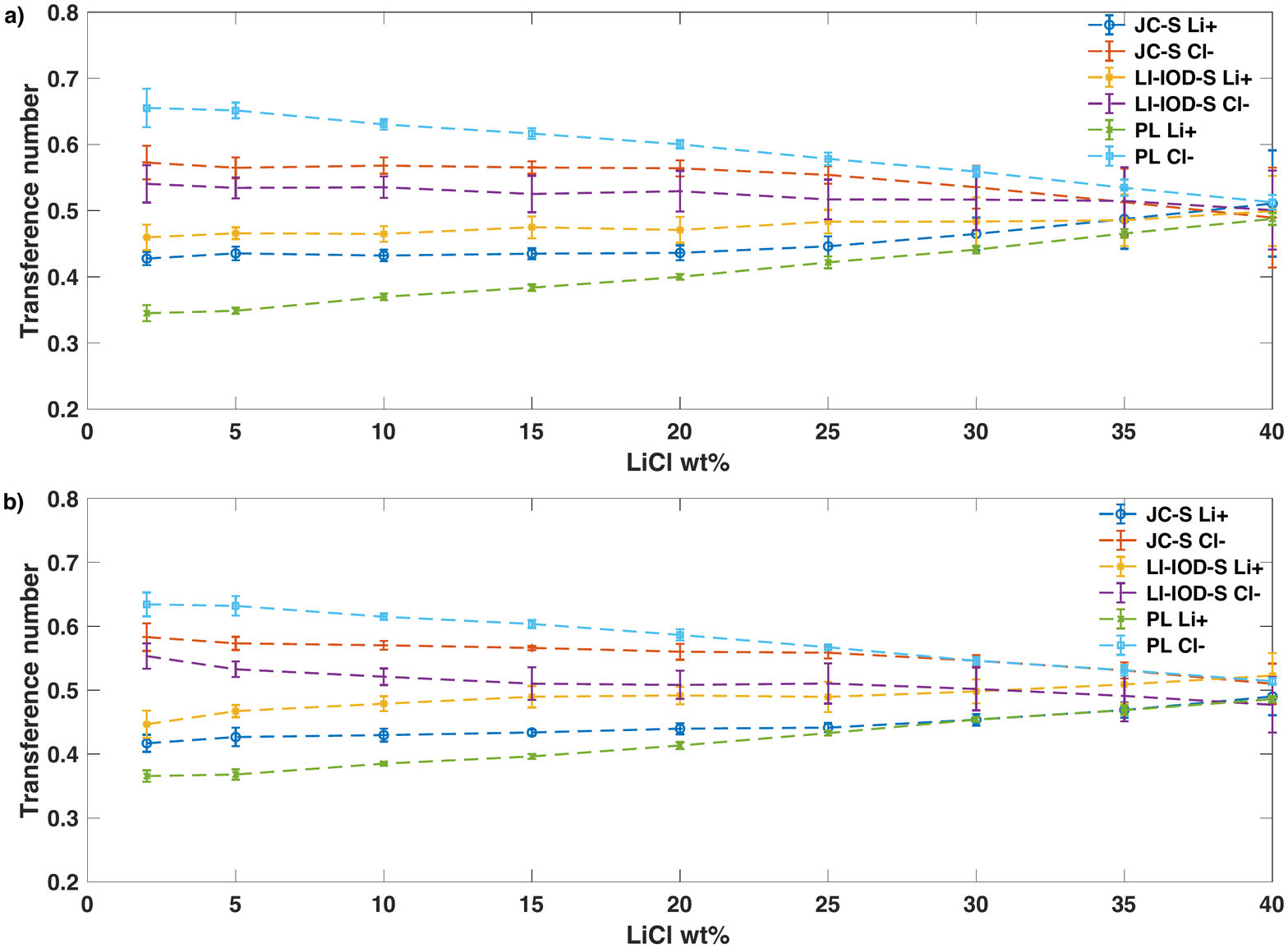}
	\caption{Transference numbers vs. wt \% of LiCl at 20$^\circ$C a)
          and 50$^\circ$C b).}
	\label{fig:trans}
\end{figure}

Fig.~\ref{fig:trans} shows the transference numbers of Li$^+$ and
Cl$^-$ of three models at both 20$^\circ$C and 50$^\circ$C. The chloride
ions contributes a larger fraction of the electrical current
(0.55 to 0.65) while lithium ions stand for a smaller fraction
(0.45 to 0.35). However, this gap diminishes as the concentration increase and eventually the transference numbers become similar nearly the solubility
limit. 

\subsection{Radial distribution function and ion-pairing}

The configurational distribution function
$P(\mathbf{r}^N)$ can be reduced to its two-particle version as~\cite{Chandler:1987tp}:

\begin{equation}
\rho(\mathbf{r}_1,\mathbf{r}_2)=N(N-1)\int d\mathbf{r}_3\int d\mathbf{r}_4\cdots\int
 d\mathbf{r}_N P(\mathbf{r}^N)
\end{equation}

which gives the joint probability distribution to find one particle at
position $\mathbf{r}_1$ and any other particle at $\mathbf{r}_2$. Note
that the factor $N(N-1)$ accounts for all possible pairs.

In an ideal gas, particles are uncorrelated. As a result, the
$\rho(\mathbf{r}_1,\mathbf{r}_2)$ simply equals to $N(N-1)/V^2\approx
\rho^2$ where $\rho$ is the number density. This leads to the definition of the quantity
$g(\mathbf{r}_1,\mathbf{r}_2)$ called the pair distribution function:

\begin{equation}
g(\mathbf{r}_1,\mathbf{r}_2)=\rho(\mathbf{r}_1,\mathbf{r}_2)/\rho^2
\end{equation}

This quantity reflects the density deviation from the (uncorrelated)
ideal gas. 

For isotropic fluid, this function depends upon
$|\mathbf{r}_1-\mathbf{r}_2|=r$, this makes $g(r)$ called a radial
distribution function. 

The coordination number, i.e. the number of neighbouring atoms within
first minimum of the $g(r)$ from a central atom, is define as:

\begin{equation}
\label{eq:coordnum}
n =4\pi\rho\int_0^{r_{min}} x^2g(x)dx
\end{equation}

\begin{figure}[h]
  \centering
   \includegraphics[width=\linewidth]{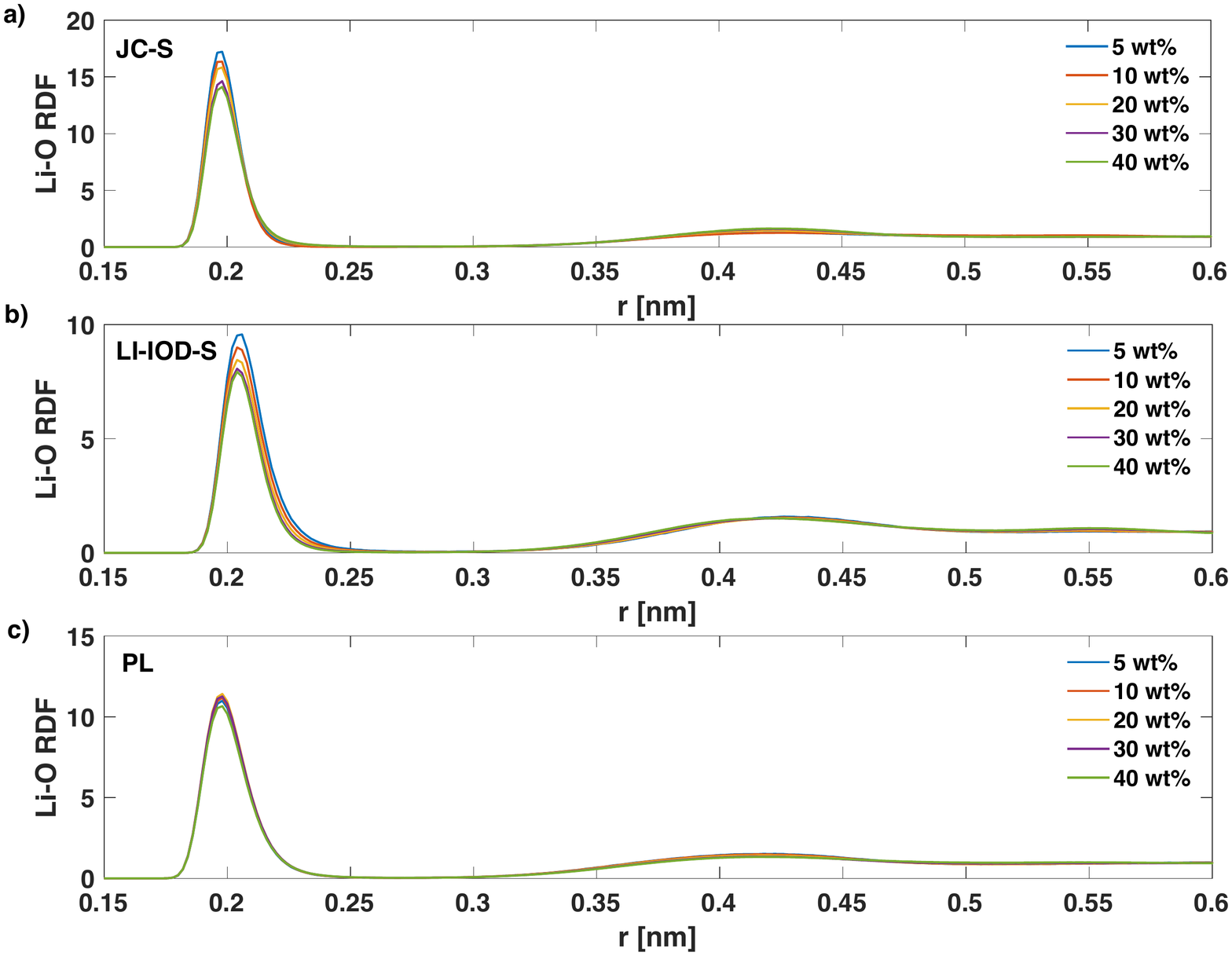}
   \caption{Radial distribution functions (RDFs) of Li$^+$-O RDF at
     5, 10, 20, 30, 40  wt\% of LiCl and the temperature of 20$^\circ$C.}
	\label{fig:rdf-li-o}
\end{figure}

\begin{figure}[h]
  \centering
   \includegraphics[width=\linewidth]{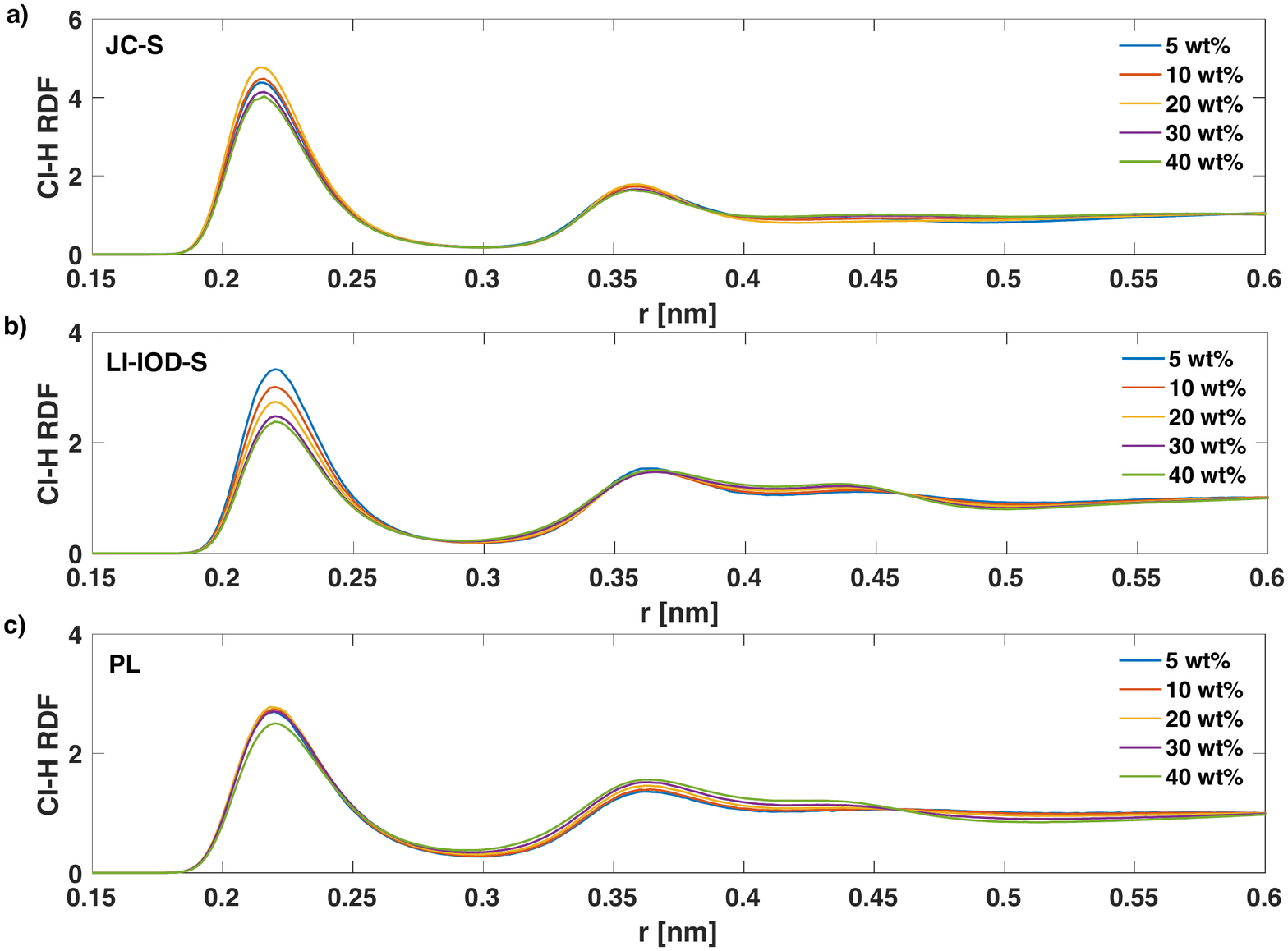}
   \caption{Radial distribution functions (RDFs) of Cl$^-$-H at
   5, 10, 20, 30, 40  wt\% of LiCl and the temperature of 20$^\circ$C.}
	\label{fig:rdf-cl-h}
\end{figure}

The first peak of $g_{\text{Li}^+-\text{O}}$ steadily
  decreases for both JC-S and LI-IOD-S ion models at 20$^\circ$C with
  increasing LiCl concentration (Fig.~\ref{fig:rdf-li-o}). Similar trend was seen for
  $g_{\text{Cl}^--\text{H}}$ with LI-IOD-S (Fig.~\ref{fig:rdf-cl-h}). This is expected, since the
  coordinating hydrogen/oxygen atoms of water molecules are gradually
  replaced by the counter-ions, see
  Table~\ref{table:cumnumrdf}. The anomaly is that
  $g_{\text{Cl}^--\text{H}}$ at 20wt\% has the highest first peak with
  JC-S. In the case of PL, the peak
  heights of  $g_{\text{Li}^+-\text{O}}$  and
  $g_{\text{Cl}^--\text{H}}$  are not much modulated by the
  concentration.

\begin{figure}[h]
  \centering
   \includegraphics[width=\linewidth]{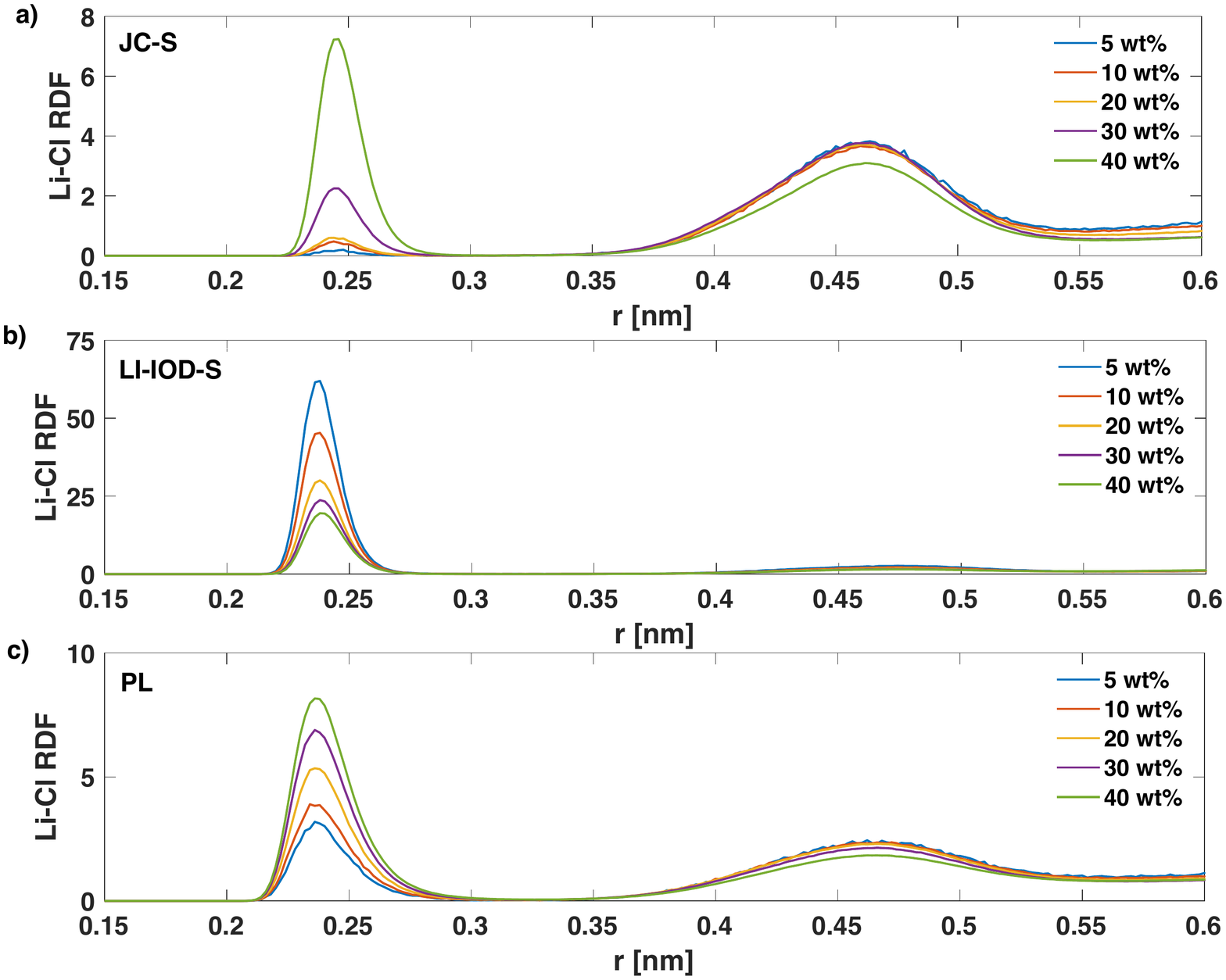}
   \caption{Radial distribution functions (RDFs) of Li$^+$-Cl$^-$  at 5, 10, 20, 30, 40  wt\% of LiCl and the temperature of 20$^\circ$C.}
	\label{fig:rdf-li-cl}
\end{figure}

Regarding the radial distribution function of Li$^+$-Cl$^-$, it goes up
  with increasing concentration (Fig.~\ref{fig:rdf-li-cl}) for JC-S
  and PL. An opposite trend was found in the case of LI-IOD-S. Despite
  that,  coordination numbers between Li$^+$ and Cl$^-$ become
  larger with the concentration for all three ion models, which
 is a sign of ion-pairing. 

We notice that the coordination number of Li$^+$-O and Cl$^-$-H in
LI-IOD-S is more sensitive to the concentration, in contrast to other two
ion models (Table~\ref{table:cumnumrdf}). This is in accord with
Fig.~\ref{fig:rdf-li-o} and Fig.~\ref{fig:rdf-cl-h}.

Although radial distribution functions at 50$^\circ$C have a similar
concentration dependence (data not shown), coordination numbers of
ion-water become
smaller in most cases as shown in Table~\ref{table:cumnumrdf} which
were expected, because hydration shells become less structured at
elevated temperature. In contrast, all three ion models show that the cation-anion coordination
number goes up with the temperature. This may be due to the fact that
the dielectric constant of liquid water decreases with the temperature and
the solvent screening is weaker accordingly. 

\begin{table}[h]
	\centering
	\caption{Coordination numbers as defined in
          Eq.~\ref{eq:coordnum} at different concentrations of LiCl. The row
          starting with the model name shows the data at 20$^\circ$C
          and the row starting with $*$ shows the corresponding data
          at 50$^\circ$C. Data at 40wt\% were rounded off to the first
          decimal to indicate a lower accuracy.}
	\label{table:cumnumrdf}
	\begin{tabular}{c c c c c c}
		\hline
		wt\% LiCl & 5 & 10 & 20 & 30 & 40 \\ \hline
          JC-S: Li$^+$-O & 4.19 & 4.16 & 4.10 & 3.87 & 3.2 \\
                     *              & 4.20 & 4.17 & 4.08 & 3.83 & 3.2  \\
		LI-IOD-S: Li$^+$-O & 3.73 & 3.30 & 2.79 & 2.34 & 2.0
          \\
          * & 3.63  &  3.2  &  2.67  &  2.27  &  2.0  \\
          PL: Li$^+$-O & 3.95 & 3.88 & 3.65 & 3.28 & 2.8 \\
          *& 3.92  &  3.82  &  3.58  &  3.19  &  2.7  \\
          JC-S: Cl$^-$-H & 6.81 & 6.84 & 6.85 & 6.57 & 5.4 \\
          *&  6.65  &  6.67  &  6.64  &  6.37  &  5.3 \\
		LI-IOD-S: Cl$^-$-H & 5.88 & 5.29 & 4.50 & 3.79 & 3.3
          \\
         *  & 5.65  &  5.06  &  4.18  &  3.63  &  3.1 \\
          PL: Cl$^-$-H & 5.56 & 5.50 & 5.11 & 4.65 & 3.9 \\
          *& 5.29  &  5.27  &  4.90  &  4.32  &  3.6  \\
		JC-S: Li$^+$-Cl$^-$ & 0.00 & 0.01 & 0.03 & 0.18 &
                                                                      0.8 \\
           *&  0.01  &  0.02  &  0.07  &  0.24  &  0.8 \\
		LI-IOD-S: Li$^+$-Cl$^-$ & 0.62 & 0.96 & 1.35 & 1.75 &
                                                                     2.0 \\
          *&  0.67  &  1.01  &  1.46  &  1.81  &  2.1  \\
		PL: Li$^+$-Cl$^-$ & 0.05 & 0.12 & 0.35 & 0.71 &
                                                                1.2\\
          *&  0.07  &  0.16  &  0.4  &  0.78  &  1.3 \\
          \hline
	\end{tabular}
\end{table}

\subsection{Ion-paring contribution to the ionic conductivity}

We mentioned at the beginning of Section 3.1 that the ionic conductivity
calculated from the Nernst-Einstein equation provides an upper bound
and the actual conductivity is always smaller because of the
ion-paring (ion-ion correlation). Near the solubility limit, the ionic conductivity may be
reduced by 30\% when ion-ion correlations are taken into
considered in the calculation~\cite{Chowdhuri:2001bl}. This is similar to our estimation
based on the mean square charge displacement~\cite{DeLeeuw:1981uc}, which gives a value of
40\% for LiCl. Therefore, the overshooting of PL in the ionic
conductivity at high concentration as shown in Fig.~1 is not because
of the missing of ion-pairing contribution in Eq.~\ref{eq:nersteinstein} but likely due to the down-scaling of the charge in the model (See Table~\ref{table:potential}).  

One of the main observations in this study is that the transference
number of the chloride ion becomes similar to that of the lithium
ion. This is in accord to the tracer diffusion measurement reported in
the literature~\cite{Tanaka:1987kg}. The standard explanation for this phenomenon is that
lithium and chloride ions pair up at high concentration and move
together in a concerted manner.

\begin{figure}[h]
  \centering
   \includegraphics[width=\linewidth]{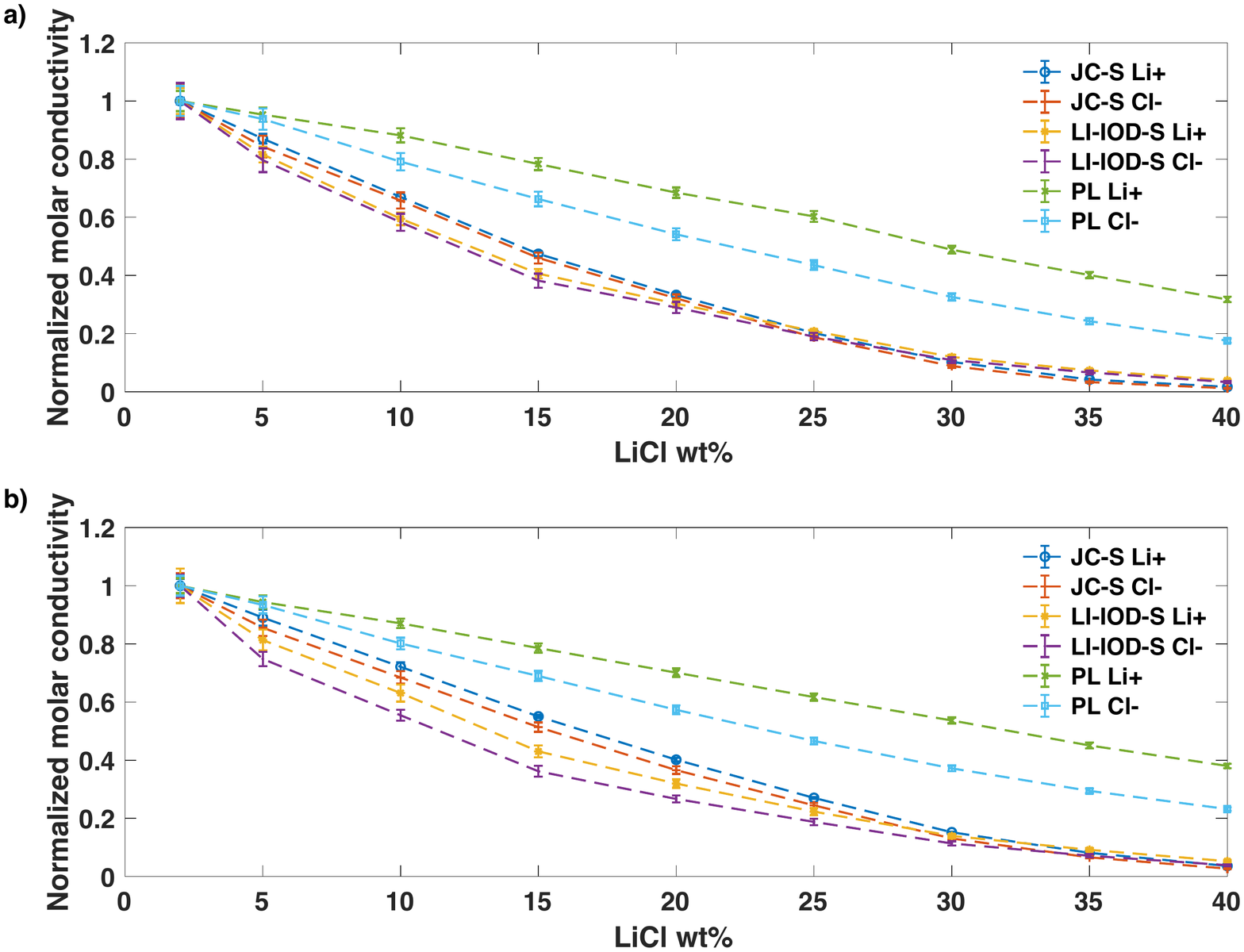}
   \caption{The normalized molar conductivities of Li$^+$ and Cl$^-$
     vs. wt\% of LiCl at 20$^\circ$C a) and 50$^\circ$C b).}
   \label{fig:relative-cond}
\end{figure}

Instead, we notice that the relative reduction of the Cl$^-$
conductivity with the increase of the concentration can be notably larger than
that of the Li$^+$ conductivity in the case of PL (Fig.~\ref{fig:relative-cond}).
. This observation is interesting because the PL model was adjusted to take into account the missing electronic
polarization. In addition, we noticed that the difference in the concentration dependence between Li$^+$ and Cl$^-$ becomes more apparent at 50$^o$C because the dielectric constant goes down. Thus, it would be of interest to investigate the effect of polarization on the ion-specific concentration dependence of mobility.

The implication of this observation is twofold.  Firstly, since the molar conductivity of the
chloride ion at infinite dilution is larger than that of the lithium
ion, therefore the ion-specific concentration dependence could lead to a crossover between cation
transference numbers and anion transference number even without
considering the ion-pairing. Secondly, the standard Debye-Onsager
theory may need be expanded in order to consider ion-specific concentration dependence for
mobility of Li$^+$ and Cl$^-$.

\section {Conclusion}

In this work, we carried out both experimental measurements and molecular
dynamics simulations of the ionic conductivity in LiCl solutions.  Ionic
conductivities were reported at both 20$^\circ$C and 50$^\circ$C from
experiments and compared to those calculated from molecular
dynamics simulations using three different ion models. In addition to provide reference data for future
force-field developments, the main finding of our study is that
transference numbers of Li$^+$ and Cl$^-$ become similar at high
concentration. This phenomenon is independent of the force fields
employed in the simulation and may be due to the ion-specific
concentration dependence of mobility.



\section{Acknowledgments}
C.Z. thanks Uppsala University for a start-up grant. Funding from the
Swedish National Strategic e-Science program eSSENCE is also gratefully acknowledged.






\end{document}